\documentclass[aps,prd,showpacs,superscriptaddress,amsmath,amsfont,amssymb,preprint,nofootinbib]{revtex4}
\usepackage{color,graphicx,epsfig}
\usepackage{ifpdf}
\usepackage{amsmath}
\usepackage{bm}
\usepackage{color}
\usepackage[english]{babel}
\usepackage{graphicx}
\usepackage{amsfonts}
\usepackage{amssymb}
\usepackage{hyperref}
\usepackage{enumerate}

\definecolor{nicered}{rgb}{0.7,0.1,0.1}
\definecolor{nicegreen}{rgb}{0.1,0.5,0.1}
\hypersetup{colorlinks,citecolor= nicegreen,linkcolor= nicered}

\newcommand{\bea}{\begin{eqnarray}}
\newcommand{\eea}{\end{eqnarray}}

\definecolor{Red}{rgb}{1.,0.,0.}

\def\gsim{{~\raise.15em\hbox{$>$}\kern-.85em
          \lower.35em\hbox{$\sim$}~}}
\def\lsim{{~\raise.15em\hbox{$<$}\kern-.85em
          \lower.35em\hbox{$\sim$}~}}

\definecolor{nicered}{rgb}{0.7,0.1,0.1}
\definecolor{nicegreen}{rgb}{0.1,0.5,0.1}
\hypersetup{colorlinks,citecolor= nicegreen,linkcolor= nicered}

\newcommand{\beq}{\begin{equation}}
\newcommand{\eeq}{\end{equation}}

\def\mysection#1{{{\bf #1}.~}}
\def\OMIT#1{}

\begin{document}

\def\Cincy{Department of Physics, University of Cincinnati, Cincinnati, Ohio 45221, USA}
\def\MIT{Center for Theoretical Physics, Massachusetts Institute of Technology, Cambridge, MA 02139, USA}

\preprint{MIT-CTP/4817}

\title{
Light quark Yukawa couplings from Higgs kinematics 
}

\author{Yotam Soreq} 
\email[Electronic address:]{soreqy@mit.edu} 
\affiliation{\MIT}

\author{Hua Xing Zhu} 
\email[Electronic address:]{zhuhx@mit.edu} 
\affiliation{\MIT}

\author{Jure Zupan} 
\email[Electronic address:]{zupanje@ucmail.uc.edu} 
\affiliation{\Cincy}

\begin{abstract}
We show that the normalized Higgs production $p_T$ and $y_h$ distributions are sensitive probes of Higgs couplings to light quarks. For up and/or down quark Yukawa couplings comparable to the SM $b$ quark Yukawa the $\bar u u$ or $\bar d d$ fusion production of the Higgs could lead to appreciable softer $p_T$ distribution than in the SM. The rapidity distribution, on the other hand, becomes more forward.  We find that, owing partially to a downward fluctuation, one can derive competitive bounds on the two couplings using ATLAS measurements of  normalized $p_T$ distribution at 8\,TeV. With 300 fb${}^{-1}$ at 13\,TeV LHC one could establish flavor non-universality of the Yukawa couplings in the down sector.
\end{abstract}

\maketitle

\section{Introduction} \label{sec:introduction}
The Higgs mechanism in the Standard Model~(SM) has a dual role -- it breaks the electroweak gauge symmetry and endows the SM charged fermions with a nonzero mass. Measurements of the Higgs production and decays by  ATLAS and CMS show that the Higgs is the dominant source of EWSB~\cite{Khachatryan:2016vau}. The Higgs mechanism also predicts that the Higgs  couplings to the SM charged fermions, $y_f^{\rm SM}$,  are proportional to their masses, $m_f$,
\beq\label{eq:SMYukawa}
y_f^{\rm SM}=\sqrt 2 m_f/v,
\eeq
with $v=246$ GeV. In the SM the Yukawa couplings are thus predicted to be very hierarchical. The prediction \eqref{eq:SMYukawa} can be distilled into four distinct questions \cite{Nir:2016zkd,Dery:2013rta}: i)~are the Yukawa couplings flavor diagonal, ii)~are the Yukawa couplings real, iii)~are diagonal Yukawa couplings proportional to the corresponding fermion masses $y_f\propto  m_f$, iv)~is the proportionality constant $\sqrt 2/v$? Given the current experimental bounds, see below, it is still possible that light fermions Yukawa are larger than the SM predictions~\cite{Giudice:2008uua,Bauer:2015fxa,Bauer:2015kzy,Bishara:2015cha,Dery:2014kxa} or much smaller due to non SM masses generation mechanism of the light fermions~\cite{Ghosh:2015gpa}. 

Experimentally, we only have evidence that the Higgs couples to the 3rd generation charged fermions \cite{Khachatryan:2016vau}. This means that the couplings to the 3rd generation charged fermions follow the hierarchical pattern \eqref{eq:SMYukawa} within errors 
from the global fits that are about ${\mathcal O}(20\%)$ (though with some preference for increased top Yukawa and decreased bottom Yukawa). 
A related question is whether Higgs couplings to the 1st and 2nd generations are smaller than the couplings to the 3rd generation. This is already established for charged leptons~\cite{Aad:2015gba,Khachatryan:2014aep} and up-type quarks \cite{Perez:2015aoa},  while flavor universal Yukawa couplings are still allowed for down quarks (for future projections see~\cite{Altmannshofer:2015qra,Perez:2015lra}), 
\beq
\frac{y_{e(\mu)}^{\rm exp}}{y_\tau^{\rm exp}}<0.22(0.28) \, , \qquad 
\frac{y_{u(c)}^{\rm exp}}{y_t^{\rm exp}}< 0.036 \, , \qquad
 \frac{y_{d(s)}^{\rm exp}}{y_b^{\rm exp}}<5.6 \, .
\eeq
The bounds on lepton Yukawa couplings come from direct searches, while the bound on light quark Yukawa couplings come from a global fit (including electroweak precision data) varying all the Higgs couplings. Significantly looser but model independent bounds on $y_c$ from a recast of $h\to b\bar b$ searches~\cite{Perez:2015aoa} or from measurements of total decay width~\cite{Perez:2015aoa,Zhou:2015wra} also show $y_c<y_t$.
 
In this manuscript we show that the indirect sensitivity to the light quark Yukawa couplings can be improved by considering normalized $d\sigma_h/d p_{T}$ or $d\sigma_h /d y_h$ distributions for the Higgs production. Higgs $p_T$ distributions have been considered before as a way to constrain new particles running in the $ggh$ loop \cite{Arnesen:2008fb,Biekotter:2016ecg,Brehmer:2015rna,Dawson:2015gka,Schlaffer:2014osa,Grojean:2013nya,Langenegger:2015lra,Bramante:2014hua,Buschmann:2014twa,Azatov:2013xha,Banfi:2013yoa,Buschmann:2014sia}.
\footnote{A related question on how to characterize the properties of a heavy resonance using kinematical distributions of its decay products and distributions in the number of jets was recently discussed in \cite{Gao:2015igz,Csaki:2016raa,Bernon:2016dow,Ebert:2016idf,Carmona:2016qgo,Harland-Lang:2016vzm}.}
  In the case of enhanced light quark Yukawa couplings the $h+j$ diagrams are due to the $q\bar q$, $qg$, and $\bar q g$ initial partonic states (the effects due to the
   inclusion of  $u,d,s$ quarks in the $ggh$ loops are logarithmically enhanced, 
but still small \cite{Melnikov:2016emg}). Since these give different $d\sigma_h/d p_{T}$ or $d\sigma_h /d y_h$ distributions than the gluon fusion initiated Higgs production, the two production mechanisms can be experimentally distinguished.
  
The first measurements of the $d\sigma/d p_T$ or $d\sigma /d y_h$ differential distributions were already performed by ATLAS~\cite{Aad:2014lwa,Aad:2014tca,Aad:2015lha,Aad:2016lvc} and CMS~\cite{Khachatryan:2015rxa,Khachatryan:2016vnn} using the Run~1 dataset. We use these to demonstrate our method and set indirect upper bounds on the up and down Yukawa couplings. The present $\mathcal{O}(30-70)$\% error is expected to be improved in the 13\,TeV LHC. As we show below at 13 TeV the LHC on can establish indirectly whether or not Higgs couples hierarchically to down-type quarks.  These can be compared with the prospects for measuring light quark Yukawa couplings in exclusive production, $h+J/\psi$,  $h+\phi$, $h+\rho$, $h+\omega$~\cite{Bodwin:2013gca,Kagan:2014ila,Koenig:2015pha}, which appear to be even more challenging experimentally and require larger statistical samples \cite{Perez:2015lra}. Other non-accelerator based suggestions for potentially probing light quark Yukawas can be found in Refs. \cite{Delaunay:2016brc,Bishara:2015cha,Celis:2014roa}. 

The paper is organized as follows. In Section  \ref{sec:distribs} we discuss the state of the art theoretical predictions of the normalized $p_T$ and $y_h$ distributions, and the sensitivity to light quark Yukawas. The present constraints and future projections are given in Section \ref{sec:constraints}, while Conclusions are collected in Section \ref{sec:Conclusions}.

\section{Light Yukawa couplings from Higgs distributions} \label{sec:distribs}

In the rest of the paper we normalize the light quark Yukawa couplings to the SM $b$-quark one, and introduce \cite{Kagan:2014ila}
\beq\label{eq:kappaq}
\bar \kappa_q = \frac{y_q^{\rm exp}}{y_b^{\rm SM}},
\eeq
{ where the Yukawa couplings are evaluated at $\mu=m_h$.} 
Establishing the hierarchy among down-type quark Yukawas thus requires showing that $\bar \kappa_d/\bar \kappa_b <1$ and/or $\bar \kappa_s/\bar \kappa_b<1$. 
{ 
Note that $\bar \kappa_{s(d)}=1$ requires a large enhancement of the Yukawa coupling over its standard model value by a factor of $\simeq 50 ~(\simeq 10^3)$.
} 
The present experimental bounds are $\bar \kappa_u <0.98 (1.3)$, $\bar \kappa_d < 0.93 (1.4)$, $\bar \kappa_s<0.70 (1.4)$, obtained from a global fit to Higgs production (including electroweak precision data) varying only the Yukawa coupling in question (or all of the Higgs couplings) \cite{Kagan:2014ila}. The sensitivity can be improved if one uses inclusive cross section at different collision energies \cite{Gao:2013nga}.

\begin{figure}[!t]
\centering
\includegraphics[width=0.47\textwidth]{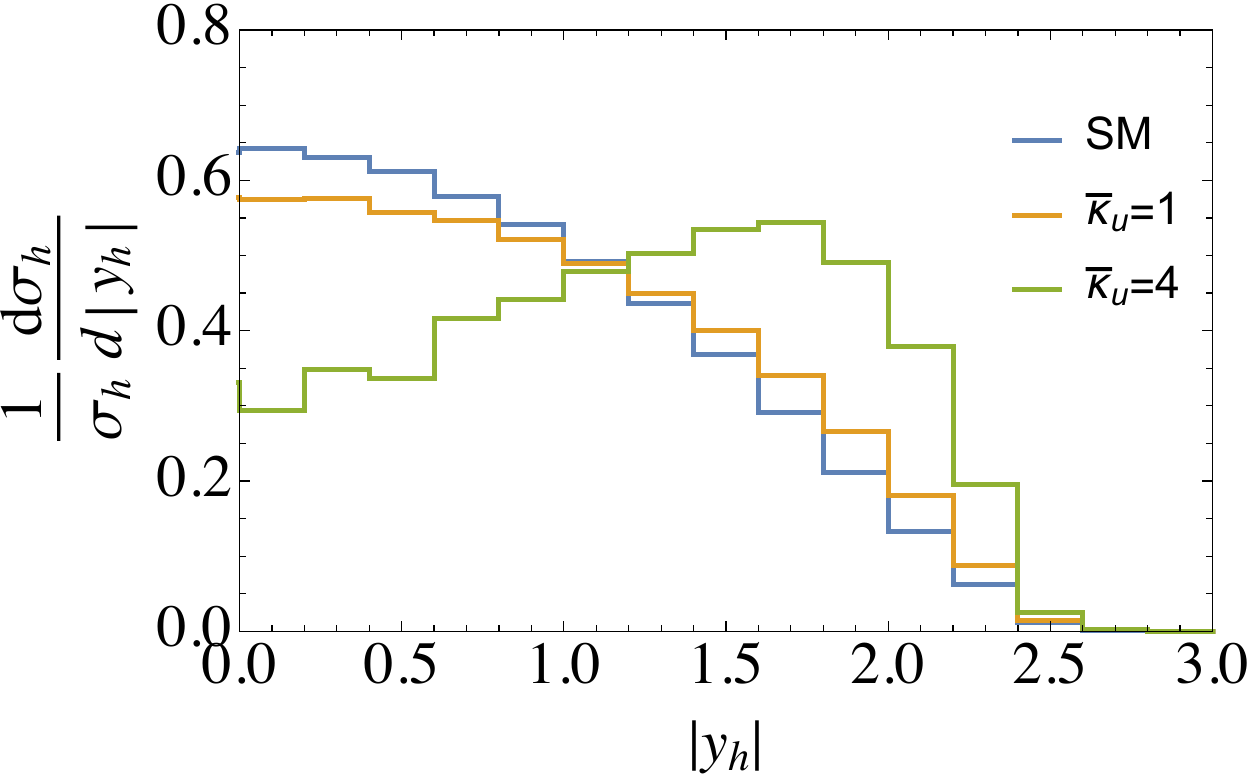}~~~~
\includegraphics[width=0.47\textwidth]{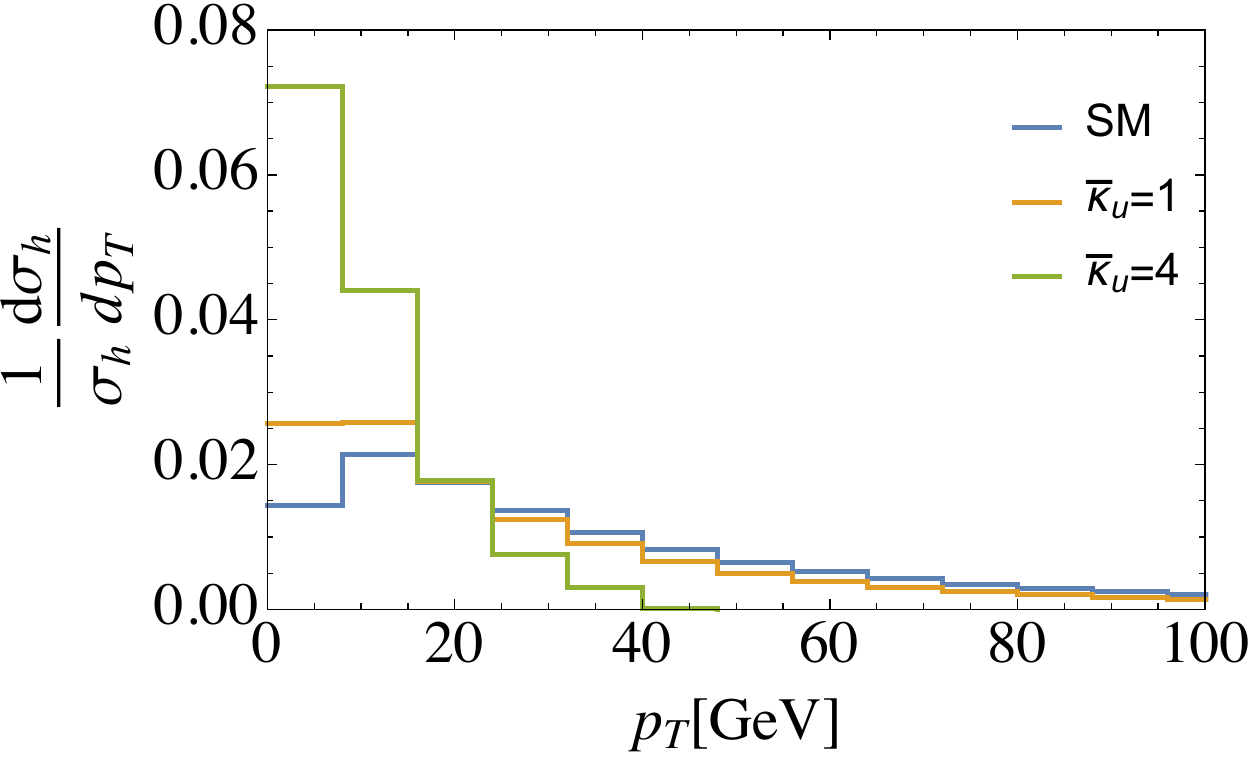}
\caption{The $1/\sigma_h\cdot d\sigma_h/dy_h$~(left) and $1/\sigma_h\cdot d\sigma_h/dp_T$~(right) normalized distributions 
{ 
at $\sqrt s=13$~TeV collision energy
} for several values of up quark Yukawa couplings, $\bar \kappa_u=0$ (SM, blue), $\bar\kappa_u=1$ (orange),  $\bar \kappa_u=4$ (green). }
\label{fig:etadistrib}
\end{figure}

In order to improve these bounds we exploit the fact that the Higgs rapidity and $p_T$ distributions provide higher sensitivity to the light quark Yukawa couplings than merely measuring the inclusive Higgs cross section. In the SM the leading order~(LO) inclusive Higgs production is through gluon fusion. This is dominated by a threshold production where both gluons carry roughly equal partonic $x$. The resulting $d \sigma_h/dy_h$ thus peaks at $y_h=0$. The distribution would change, however, if one were to increase the Yukawa coupling to $u$-quarks, such that the LO Higgs production would be due to $u\bar u$ fusion. Since $u$ is a valence quark the $u\bar u$ fusion is asymmetric, with $u$ quark on average carrying larger partonic $x$ than the $\bar u$  sea quark. The Higgs production would therefore peak in the forward direction. This is illustrated in Fig.~\ref{fig:etadistrib}~(left) where 
the rapidity distribution is plotted for different values of the up Yukawa, $\bar \kappa_u=y_u/y_b^{\rm SM}=1.0$~(orange line) and 4.0~(green line), and for setting up Yukawa couplings to zero, $\bar \kappa_u=0$ (blue line, denoted SM in the legend). For large values of $y_u$, a few times the SM bottom Yukawa, the Higgs production is no longer central but forward. Already for $y_u=y_b^{\rm SM}$ there is a visible reduction of the Higgs production in the central region, and an increase in the forward region.

Somewhat different considerations apply to the case of Higgs $p_T$ distribution, shown in Fig.~\ref{fig:etadistrib}~(right). Due to initial state radiations, the Higgs $p_T$ distribution exhibits a Sudakov peak at small $p_T$~\cite{Collins:1984kg} (for recent works on the resummations in this region see \cite{Monni:2016ktx,Neill:2015roa,Echevarria:2015uaa,Stewart:2013faa}). The location of the peak is sensitive to the nature of the incoming partons. For $gg$ fusion, the $p_T$ distribution peaks at about $10$ GeV, while for $u\bar{u}$ scattering, the $p_T$ distribution peaks at smaller values, at about $5$ GeV. This is because the effective radiation strength of gluon is $\alpha_s N_c$, a few times larger than the effective radiation strength of quarks, $\alpha_s(N^2_c-1)/(2 N_c)$, where $N_c = 3$. The larger effective radiation strength of gluons also leads to a harder $p_T$ spectrum. In terms of normalized $p_T$ distribution, therefore,  the $u\bar{u}$ scattering leads to a much sharper peak at lower $p_T$ compared with the $gg$ scattering.

\begin{figure}[!t]
\centering
\includegraphics[width=0.45\textwidth]{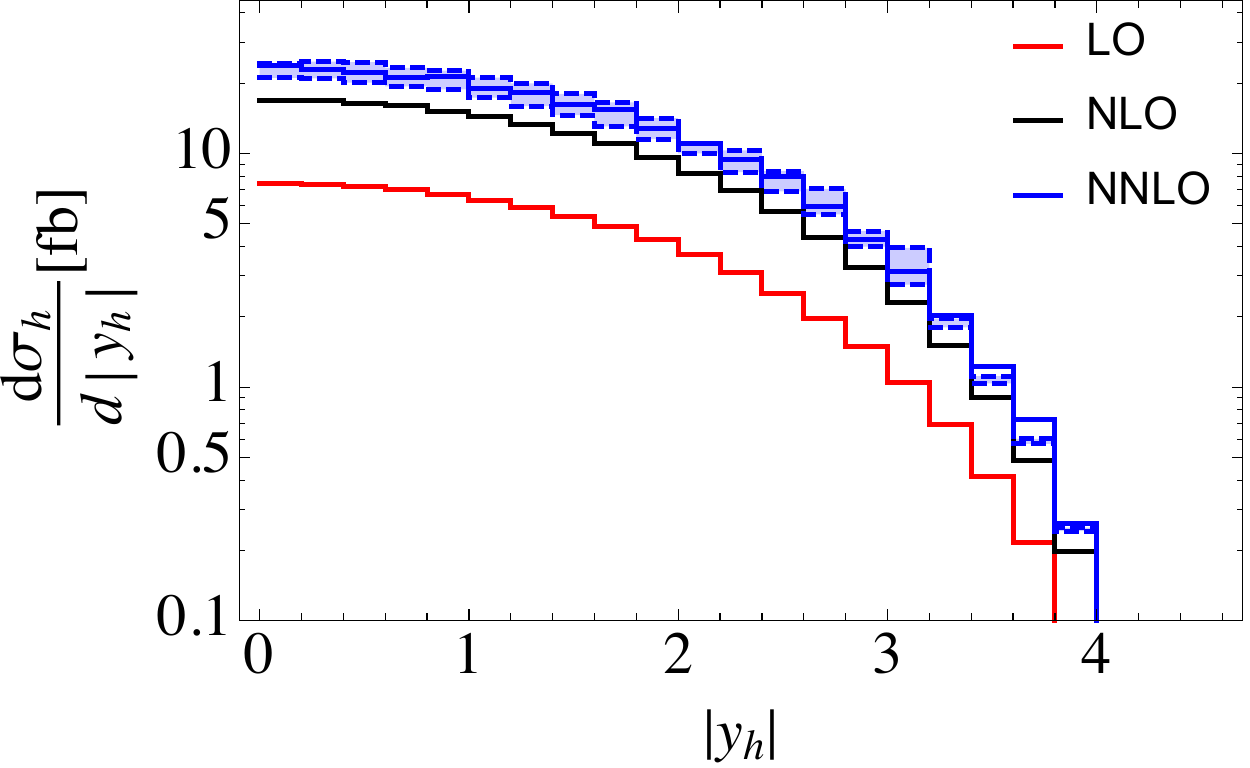}~~~~
\includegraphics[width=0.45\textwidth]{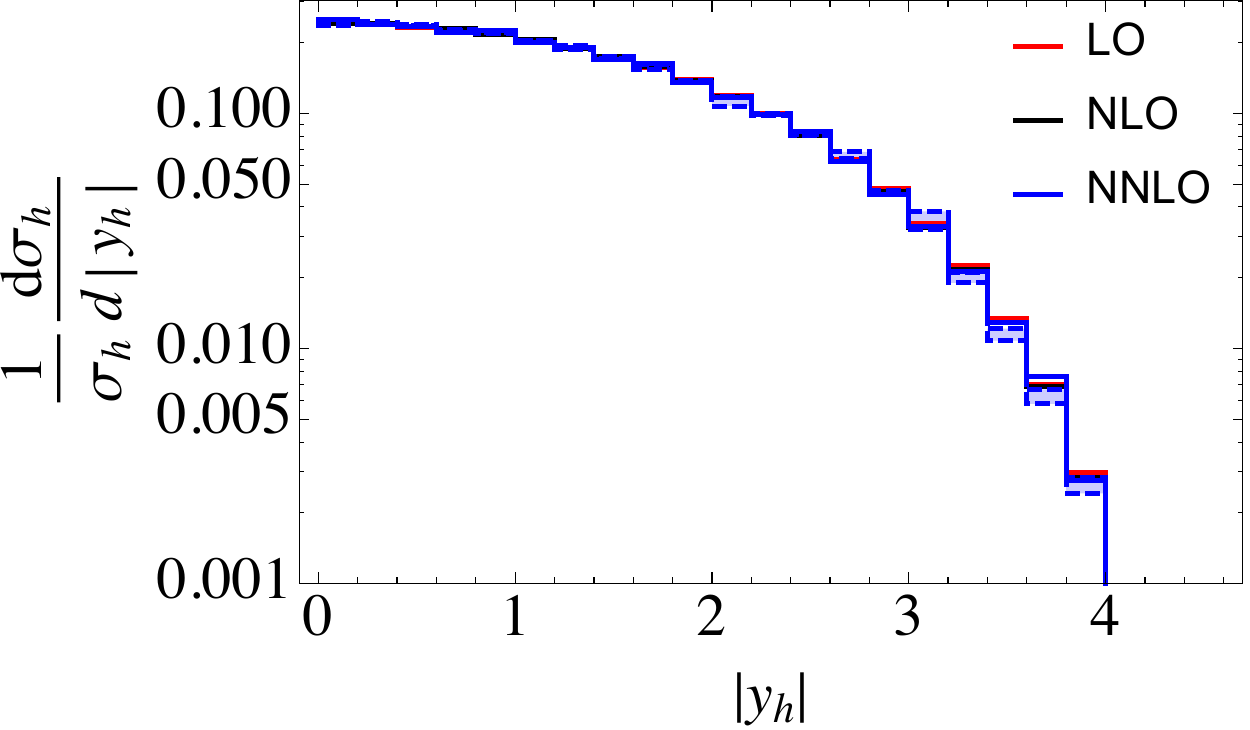} \\
\includegraphics[width=0.45\textwidth]{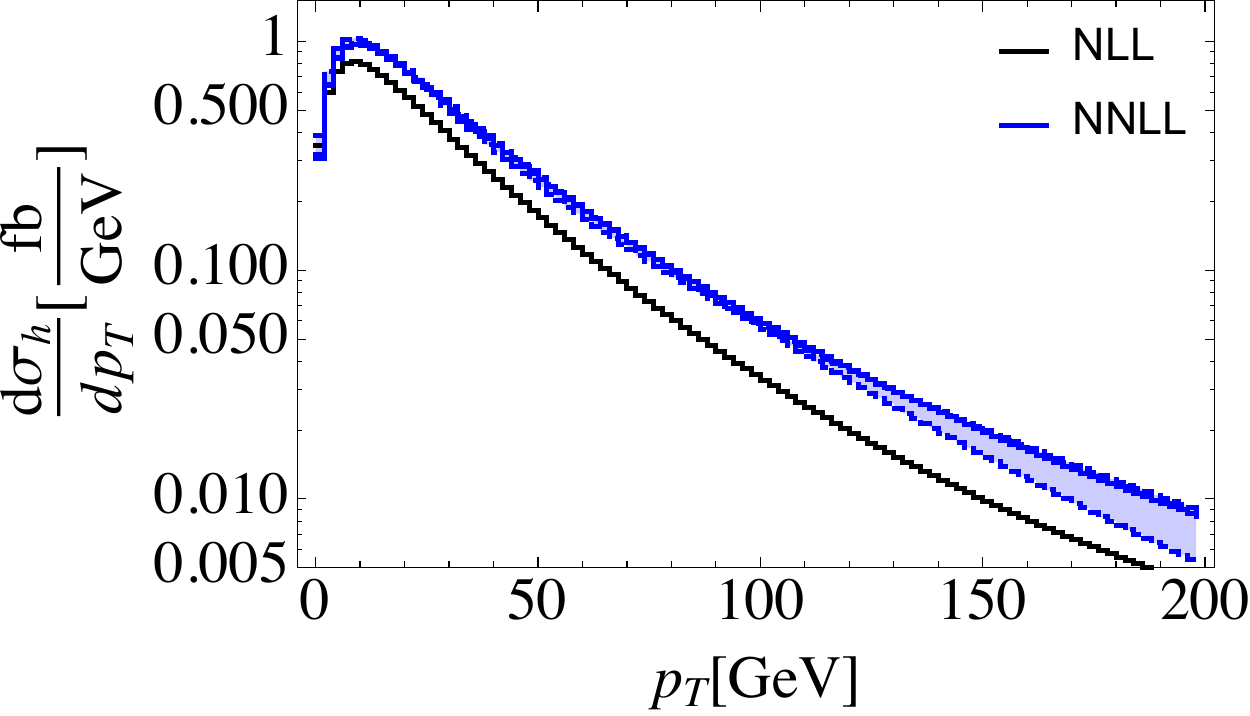}~~~~
\includegraphics[width=0.45\textwidth]{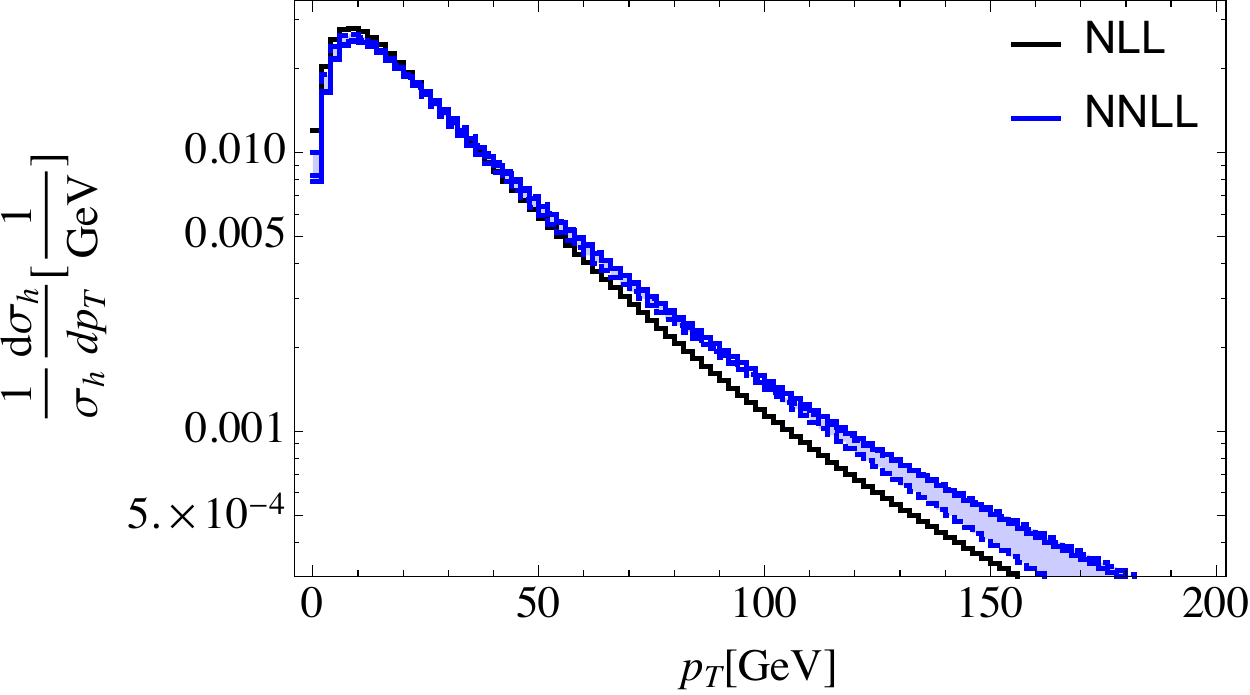} 
\caption{The upper panels show the rapidity distribution $d\sigma_h/dy_h$~(left), where the Higgs decay to $\gamma\gamma$, and the normalized rapidity distribution $1/\sigma_h\cdot d\sigma_h/dy_h$~(right) calculated at  LO, NLO and NNLO (red, black, blue lines respectively) using \texttt{HNNLO}~\cite{Catani:2007vq}, see text for details. The lower panels show NLL (black) and NNLL (blue) predictions for $d\sigma_h/dp_T$~(left) and $1/\sigma_h\cdot d\sigma_h/dp_T$~(right), obtained using \texttt{HqT2.0}\cite{Bozzi:2005wk,deFlorian:2011xf}. Blue bands denote scale dependence when varying $m_h/4<\mu<m_h$.}
\label{fig:LOvsNLOvsNNLO}
\end{figure}

Many of the theoretical errors cancel in the normalized distributions so that $1/\sigma_h\cdot d\sigma_h/dy_h$ is under much better control than the absolute value of the cross section~\cite{Anastasiou:2004xq}. This is illustrated in the top panels of Fig.~\ref{fig:LOvsNLOvsNNLO}, where we compare LO, NLO and NNLO theoretical predictions for the normalized and unnormalized $y_h$ distributions { 
at $\sqrt s=13$~TeV collision energy
}~\cite{Catani:2007vq}.  Similar cancellation of theoretical uncertainties is observed for normalized $p_T$ distribution, illustrated in the bottom panels of Fig.~\ref{fig:LOvsNLOvsNNLO}, although the reduction of theoretical uncertainties is not as dramatic as in the rapidity distribution. Normalized distribution also help reduces many of the experimental uncertainties. For un-normalized distribution, the total systematic uncertainties due to, e.g., luminosity and background estimates range from 4\% to 12\%~\cite{Aad:2015lha}. However, most of the systematic uncertainties cancel in the normalized shape distribution. The dominant experimental uncertainties for the shape of the distribution are statistical ones, ranging from 23\% to 75\% \cite{Aad:2015lha}, and can be improved with more data.

In this work we perform an initial study using the rapidity and $p_T$ distributions to constrain the light-quark Yukawa couplings. In the study we use Monte Carlo samples of events on which we impose the experimental cuts in Section \ref{sec:constraints}. We generate the parton level, $pp\to h +n$ jets, including the SM gluon fusion (the background) and $q\bar{q}$ and $q g, \bar q g$ fusion (the signal) using MadGraph\,5~\cite{Alwall:2014hca} with LO CT14 parton distribution function~(PDF)~\cite{Dulat:2015mca} and Pythia~6.4~\cite{Sjostrand:2006za} for the showering, where $q=u,d,s,c$ and $n=0,1,2$. Events of different multiplicities are matched using the MLM scheme~\cite{Mangano:2006rw}. Further re-weighting of the generated tree-level event samples is necessary because of the large $k$-factor due to QCD corrections to the Higgs production~\cite{Harlander:2002wh}.  { We re-weight the LO cross section of different jet multiplicities merged in the MLM matching scheme, to the best available theoretical predictions so far. For contributions proportional to top Yukawa coupling, which start as $gg\to h$, we use N3LO predictions~\cite{Anastasiou:2015ema,Anastasiou:2016cez}, while for contributions proportional to light quark Yukawa, which start as $q\bar{q} \to h$, we use NNLO predictions~\cite{Harlander:2003ai,Harlander:2012pb,Harlander:2015xur,Harlander:2016hcx}. We combine the two re-weighted event samples to compute the normalized differential distributions $1/\sigma_h \cdot d\sigma_h/dy_h$ and $1/\sigma_h \cdot d\sigma_h/dp_T$. Our calculation is performed in the large top quark mass limit and we ignore light-quark loop in the $gg$ fusion channel. The same procedure is applied throughout this work.
}

\begin{figure}[!t]
\centering
\includegraphics[width=0.47\textwidth]{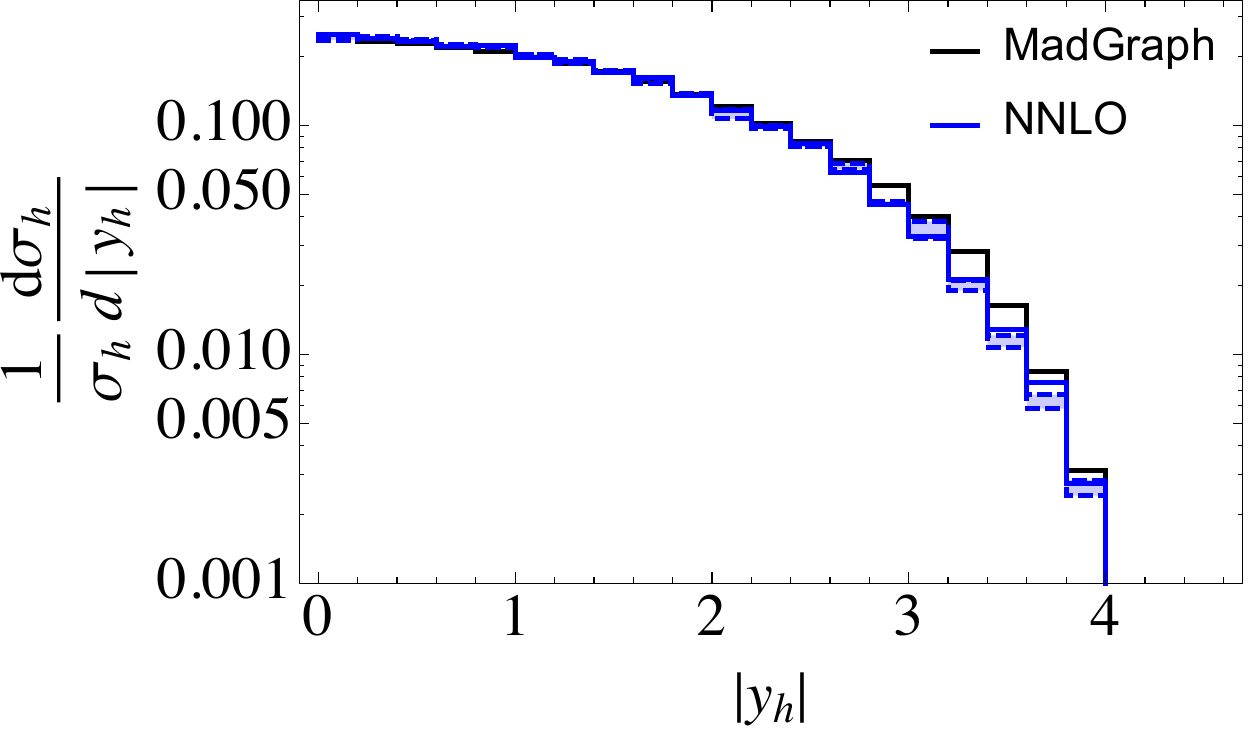}~~~~
\includegraphics[width=0.47\textwidth]{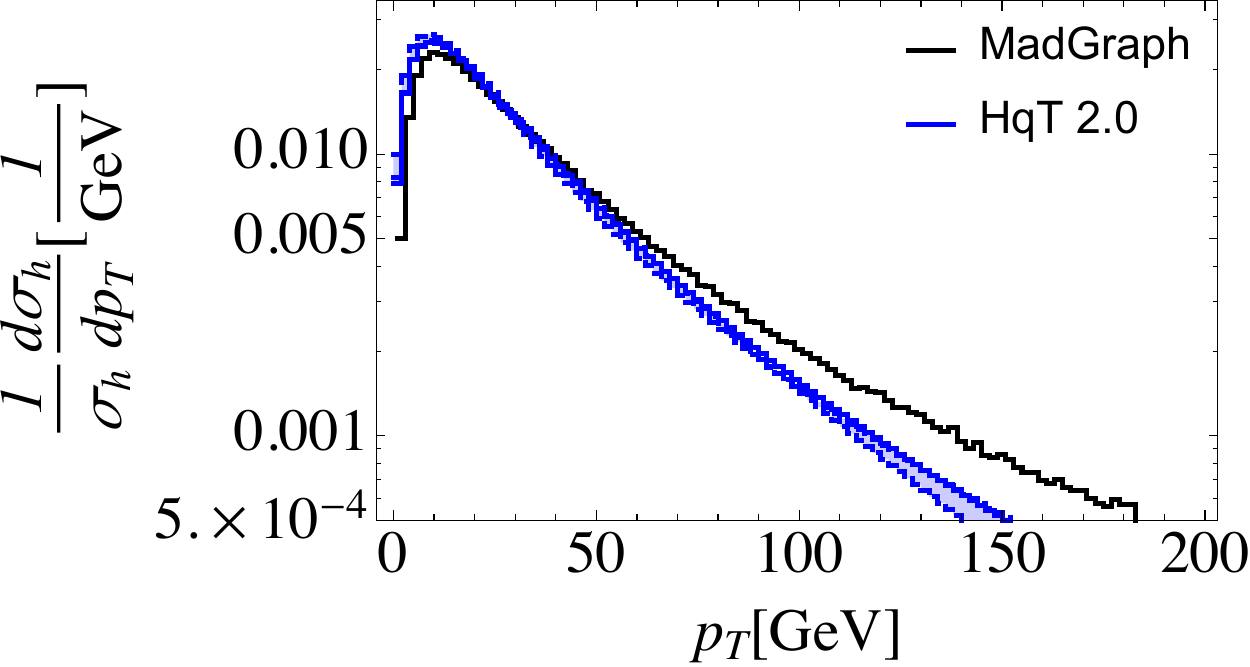}
\caption{The predictions for the $1/\sigma_h\cdot d\sigma_h/dy_h$~(left) and
  $1/\sigma_h\cdot d\sigma_h/dp_T$~(right) normalized distributions { 
at $\sqrt s=13$~TeV collision energy
}. The tree-level MadGraph\,5+Pythia is shown in solid black, while the QCD NNLO rapidity
  distribution computed using \texttt{HNNLO} code \cite{Catani:2007vq}, shown in the left panel   (the NNLO+NNLL resumed $p_T$ distributions computed using \texttt{HqT2.0} \cite{Bozzi:2005wk,deFlorian:2011xf}, shown in the right panel), are denoted by blue lines. Theoretical uncertainties, denoted by the blue bands, are estimated by varying the resummation scale between $m_h/4$ and $m_h$.}
\label{fig:etaptcheck}
\end{figure}

In Fig.~\ref{fig:etaptcheck}, we compare our tree-level MadGraph\,5+Pythia prediction for the normalized rapidity and $p_T$ distribution against the available precise QCD prediction based on NNLO~\cite{Catani:2007vq} and NNLO+NNLL calculations~\cite{deFlorian:2011xf}. We find that for the rapidity distribution the MadGraph\,5+Pythia calculation describes well the shape of the normalized distribution. Small differences at the level of ${\mathcal O}(10\%)$ are observed for the $p_T$ distribution. In the future, when experimental data become more precise, it will be useful to redo our phenomenological analysis, presented below, using more precise resummed predictions for both the signal and background.

\begin{figure}[!t]
\centering
\includegraphics[width=0.45\textwidth]{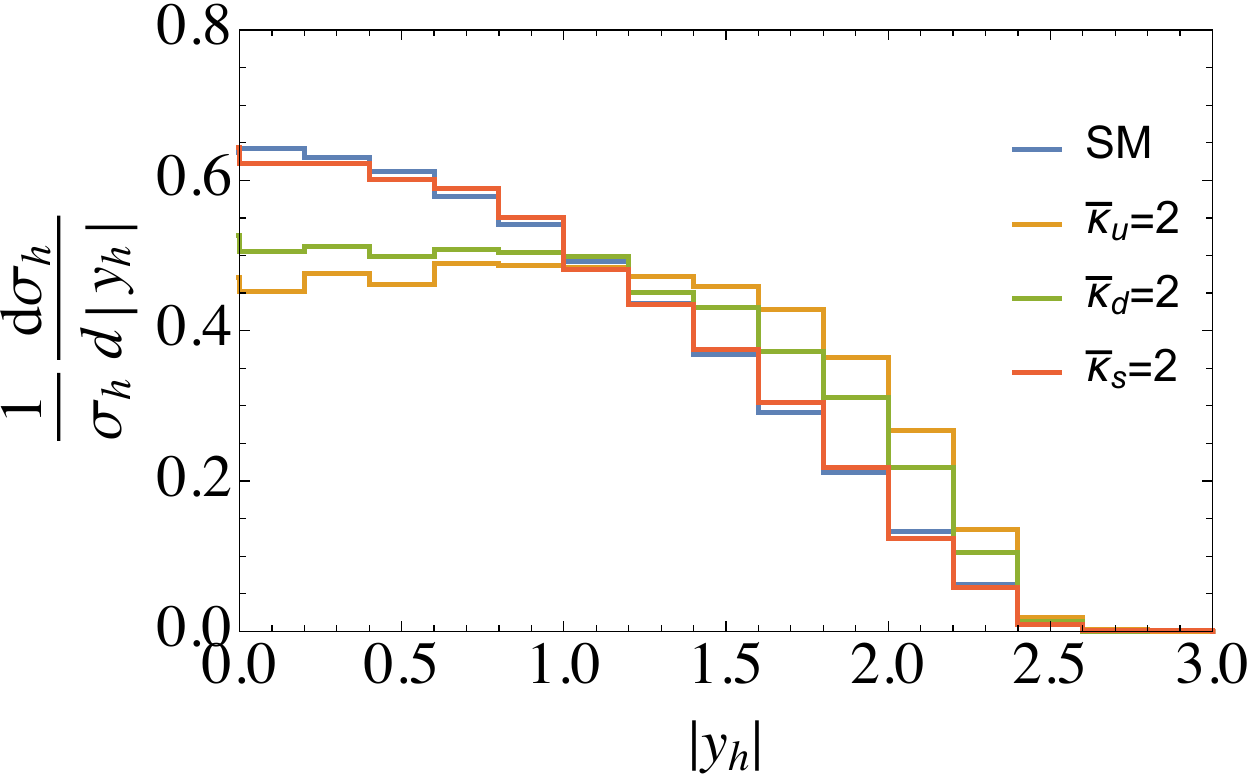}~~~~
\includegraphics[width=0.45\textwidth]{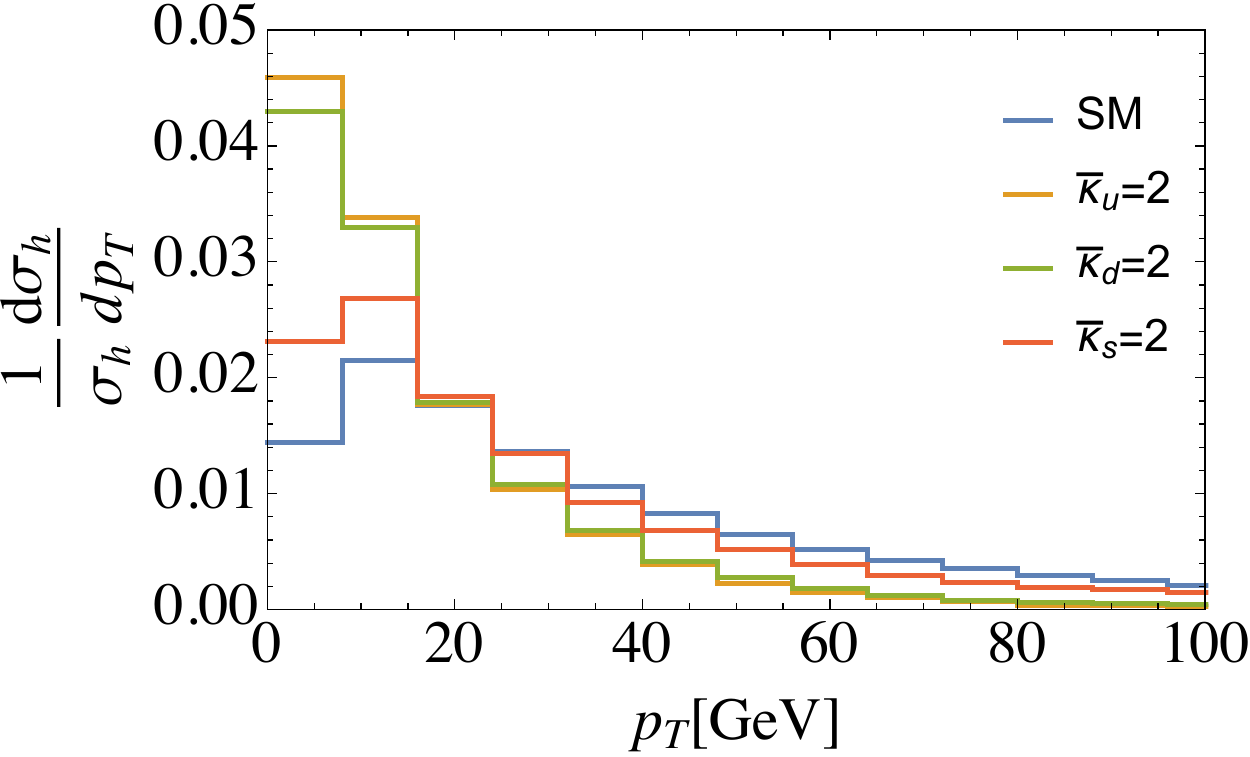}
\caption{The $1/\sigma_h\cdot d\sigma_h/dy_h$~(left) and $1/\sigma_h\cdot d\sigma_h/dp_T$~(right) when switching on up (orange),  down (green) and strange (red)  Yukawa coupling.}
\label{fig:different_yukawas}
\end{figure}

The difference between Higgs production kinematics with and without significant light quark Yukawas becomes smaller when going from $u\bar u$ fusion to $d\bar d$ fusion and to  $s\bar s$ fusion (for the same value of the Yukawa coupling in each case). In Fig.~\ref{fig:different_yukawas}, we set $y_u=y_d=y_s=2.0\times y_b^{\rm SM}$ to illustrate this point. Since $s$ is a sea quark its PDF is much closer to the gluon PDF, leading to similar Higgs $p_T$ and $y_h$ distributions in the case of pure gluon fusion and when strange Yukawa is enhanced. We therefore do not expect large improvements in the sensitivity to the strange Yukawa by considering Higgs cross section distributions compared to just using the total rates. Charm quark, on the other hand, has large enough mass that  the log enhanced contributions from the charm loop in $gg \to hj$ production can have a visible effect on the Higgs kinematical distributions~\cite{Bishara:2016jga}.  

We note that a potential direct handle on the charm Yukawa can be obtained from the $h\to c\bar{c}$ inclusive rate by using charm tagging~\cite{Delaunay:2013pja,Perez:2015aoa,Perez:2015lra,ATL-PHYS-PUB-2015-001} or from a Higgs produced in association with a $c$-jet~\cite{Brivio:2015fxa}. The sensitivity of the later may be potentially improved by considering the Higgs $p_T$ and $y_h$ distributions, or by considering a Higgs produced in association with two $c$-jets.

\section{Current constraints and future prospects} 
\label{sec:constraints}

In this section we perform a sensitivity study of the Higgs kinematical distributions as probes of the 1st generation quark Yukawas. We use normalized differential distributions, which, as argued above, have small theoretical uncertainties. In addition, the dependence on the Higgs decay properties, such as the branching ratios and total decay witdh, cancel in the measurements of $1/\sigma_h \cdot d\sigma_h/d p_T$ and $1/\sigma_h\cdot d\sigma_h/d y_h$. In other words, the normalized distributions are sensitive only to the production mechanism.

The Higgs production can differ from the SM one either by having a modified $ggh$ coupling, or by modified light quark Yukawas. The modification of the Higgs coupling to gluons can arise, for instance, from a modified top Yukawa coupling or be due to new particles running in the loop. In the normalized distribution the presence of new physics in the gluon fusion will affect the total rate and can be searched for in normalized distribution such as $1/\sigma_h \cdot d\sigma_h/d p_T$ for very hard $p_T$, larger than about $300$ GeV \cite{Schlaffer:2014osa,Grojean:2013nya,Langenegger:2015lra,Bramante:2014hua,Buschmann:2014twa,Azatov:2013xha,Banfi:2013yoa,Buschmann:2014sia}. In contrast, nonzero light quark Yukawa couplings modify the Higgs kinematics in the softer part of the $p_T$ spectrum. In our analysis we assume for simplicity that the gluon fusion contribution to the Higgs production is the SM one.

\begin{figure}[!t]
\centering
\includegraphics[width=0.49\textwidth]{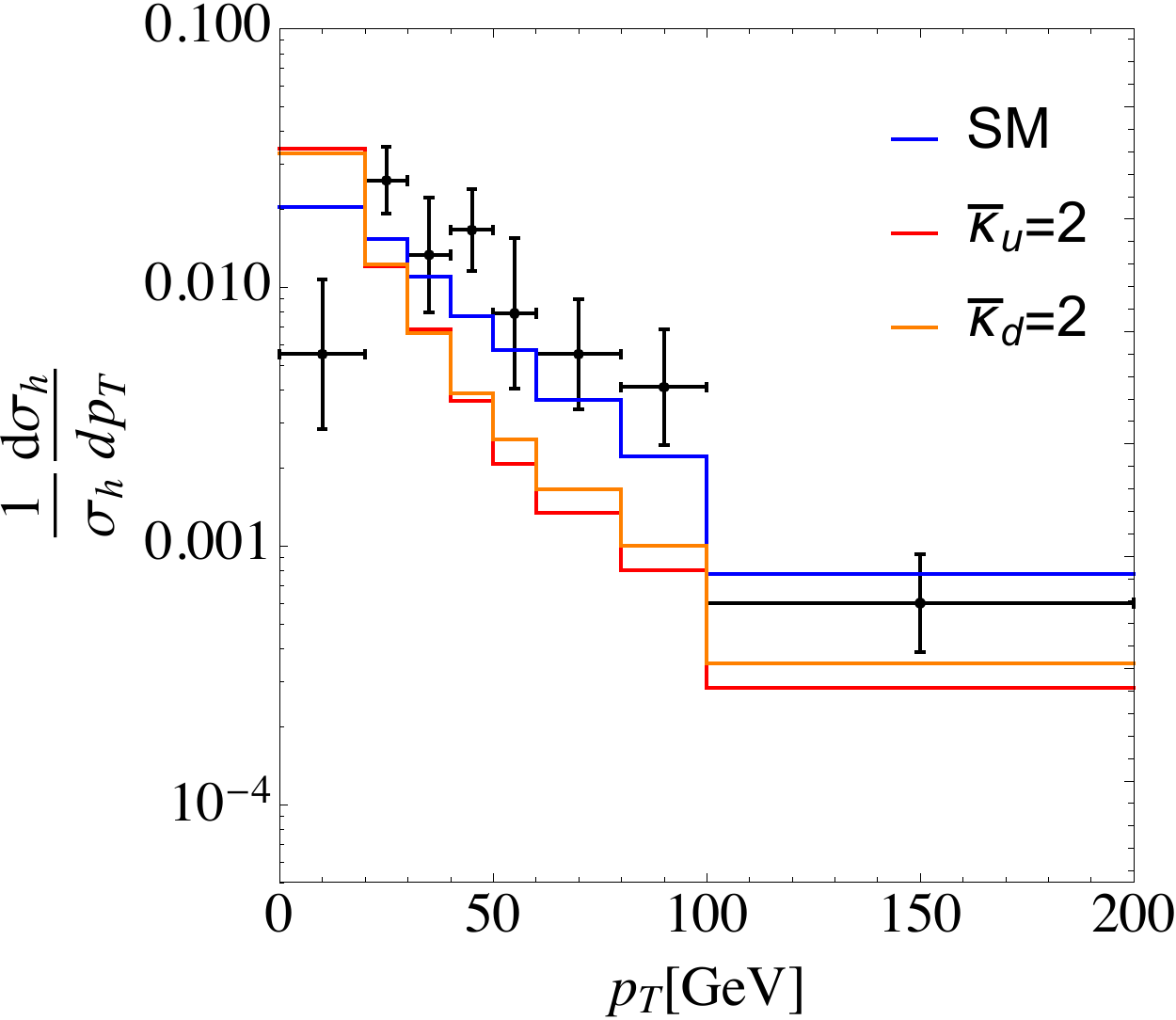}~~~~
\includegraphics[width=0.45\textwidth]{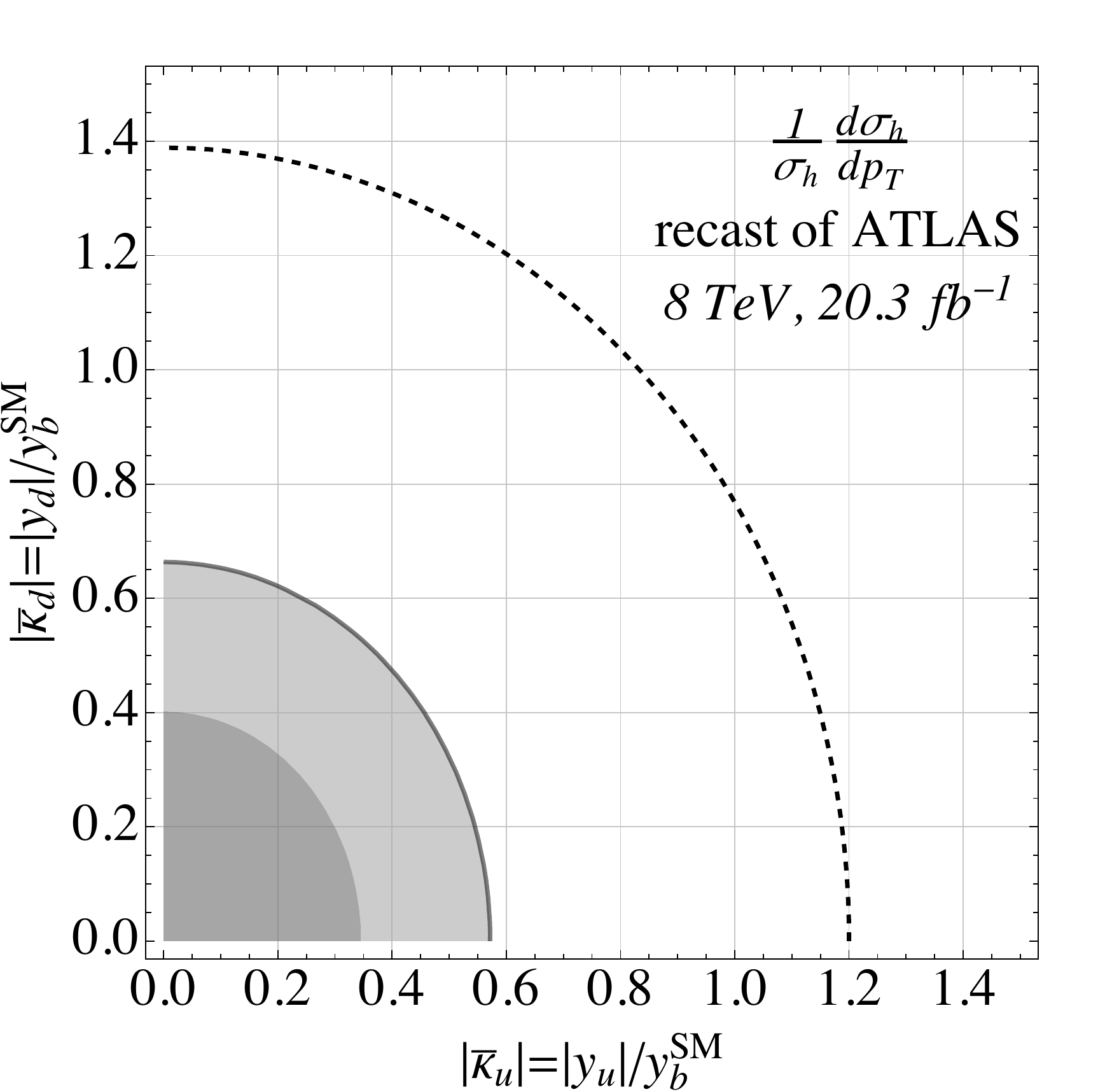}
\caption{Left: The ATLAS 8TeV measurement of the normalized  Higgs $p_T$ distribution (black)~\cite{Aad:2015lha},  and the theoretical predictions for the SM (blue), $\bar \kappa_u=2$ (red), $\bar \kappa_d=2$ (orange).  Right: The resulting $1\sigma$ ($2\sigma$) allowed regions for the up and down Yukawa are denoted by dark gray (light gray) shadings, while the dashed line denotes the $2\sigma$ expected sensitivity.}
\label{fig:8TeVpT}
\end{figure}

We use the normalized Higgs $p_T$ distribution measured by ATLAS in $h\to \gamma\gamma$ and $h\to ZZ$ channels~\cite{Aad:2015lha}, to extract the bounds on the up and down Yukawa couplings. We reconstruct the $\chi^2$ function, including the covariance matrix, from the information given in~\cite{Aad:2015lha}. The theoretical errors on the normalized distributions are smaller than the experimental ones, and can thus be neglected. The resulting  95\,\% CL regions for the up and down Yukawa are
\begin{align}\label{eq:8TeVbounds}
	\left[ \bar{\kappa}_u \right]_{{\rm 8\,TeV}, p_T} < 0.46\, , \qquad\qquad 
	\left[ \bar{\kappa}_d \right]_{{\rm 8\,TeV}, p_T} < 0.54 \, ,
\end{align}
{
where we used the Higgs $p_T$ to derive the bounds, but not the $y_H$ distributions that are less sensitive.
}
For each of the bounds above we marginalized over the remaining Yukawa coupling with the most conservative bound obtained when this is set to zero. Note that the inclusion of correlations is important. The bins are highly correlated because the distribution is normalized. The corresponding 2D contours are given in Fig.~\ref{fig:8TeVpT} (right). These bounds are stronger than the corresponding ones coming from the fits to the inclusive Higgs production cross sections, see the discussion following Eq.~\eqref{eq:kappaq}. In Fig.~\ref{fig:8TeVpT} (left) we also show the comparison  between ATLAS data~\cite{Aad:2015lha} (black), and the theoretical predictions for zero light quark Yukawas, $\bar\kappa_{u,d}=0$~(blue), and  when switching on one of them, $\bar\kappa_{u}=2$~(red) or $\bar\kappa_{d}=2$~(orange). The constraints from the Higgs rapidity distributions are at present significantly weaker.

\begin{figure}[!t]
\centering
\includegraphics[width=0.49\textwidth]{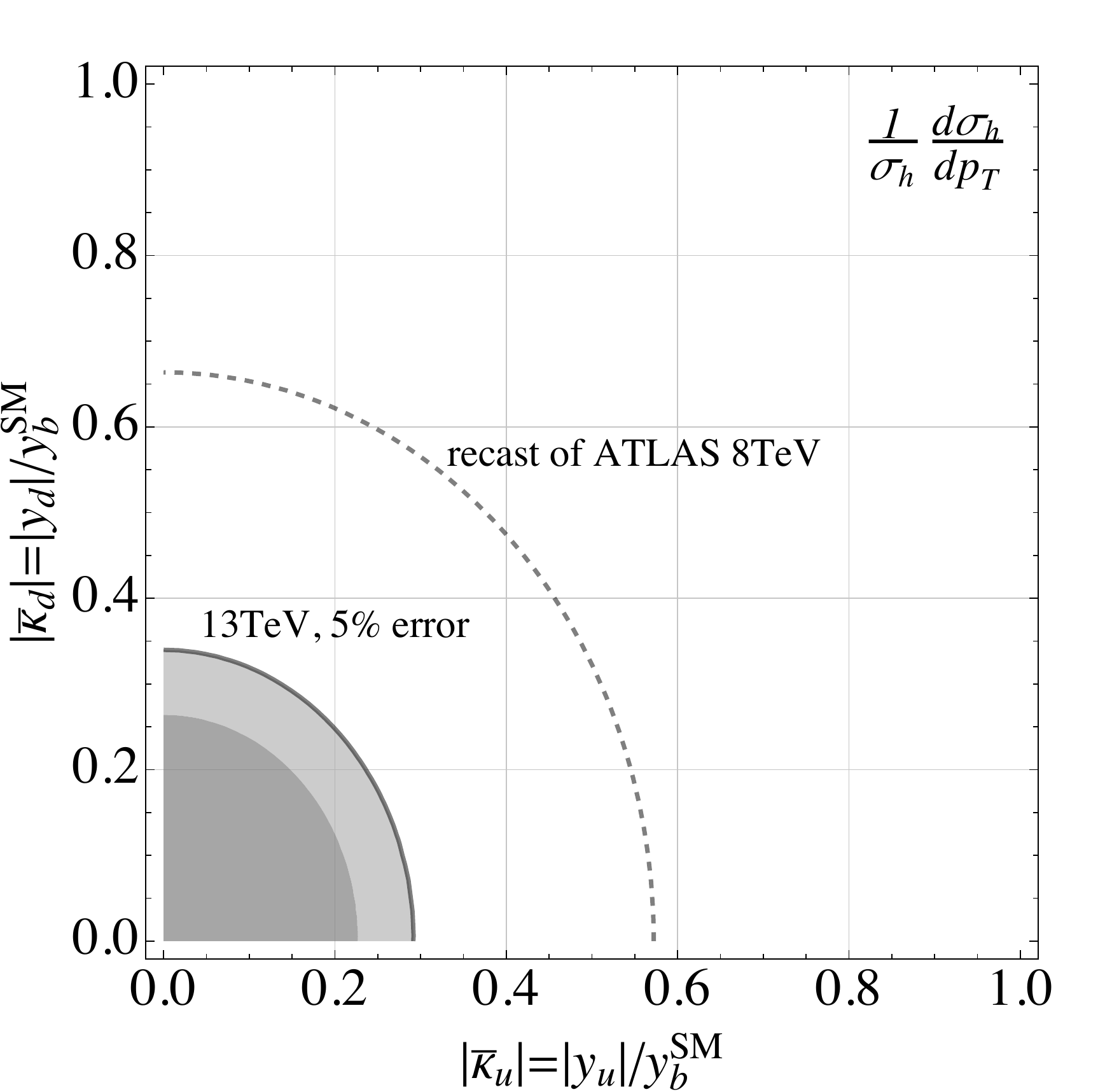}~~~~
\includegraphics[width=0.49\textwidth]{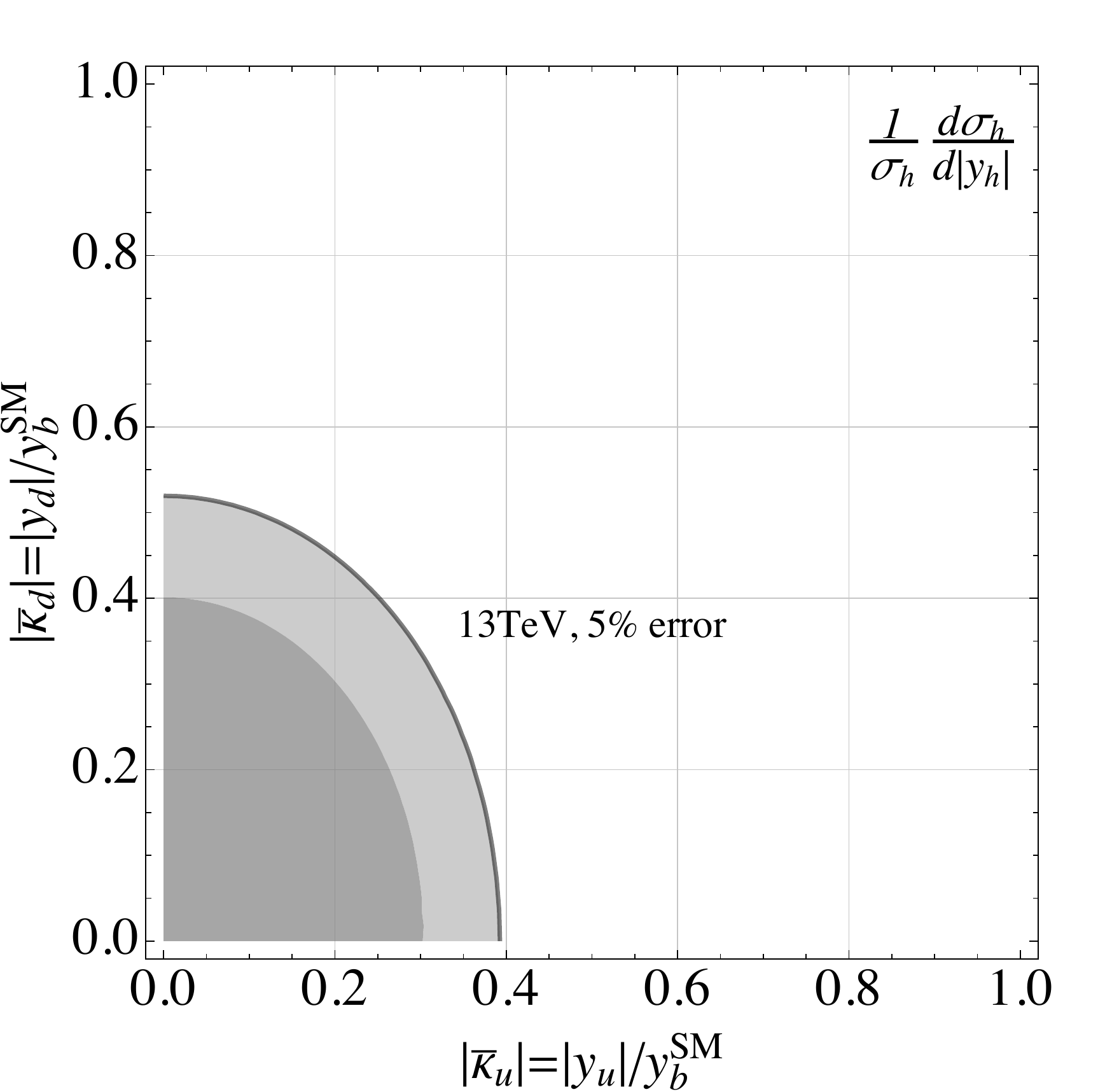}
\caption{Left: The dark\,(light) gray region is the $1\sigma$\,$(2\sigma)$ naive projection from the $p_T$ distribution for the LHC 13\,TeV assuming bin size of 10\,GeV and relative error of 5\,\% per bin. The dashed line is the current 2\,$\sigma$ bound from recast of the ATLAS 8\,TeV data. Right: The dark\,(light) gray region is the 1\,(2)$\sigma$ naive projection from the rapidity distribution for the LHC 13\,TeV assuming bin size of 0.1 and relative error of 5\,\% per bin.
 }
 \label{fig:naive13}
\end{figure}

\begin{figure}[!t]
\centering
\includegraphics[width=0.49\textwidth]{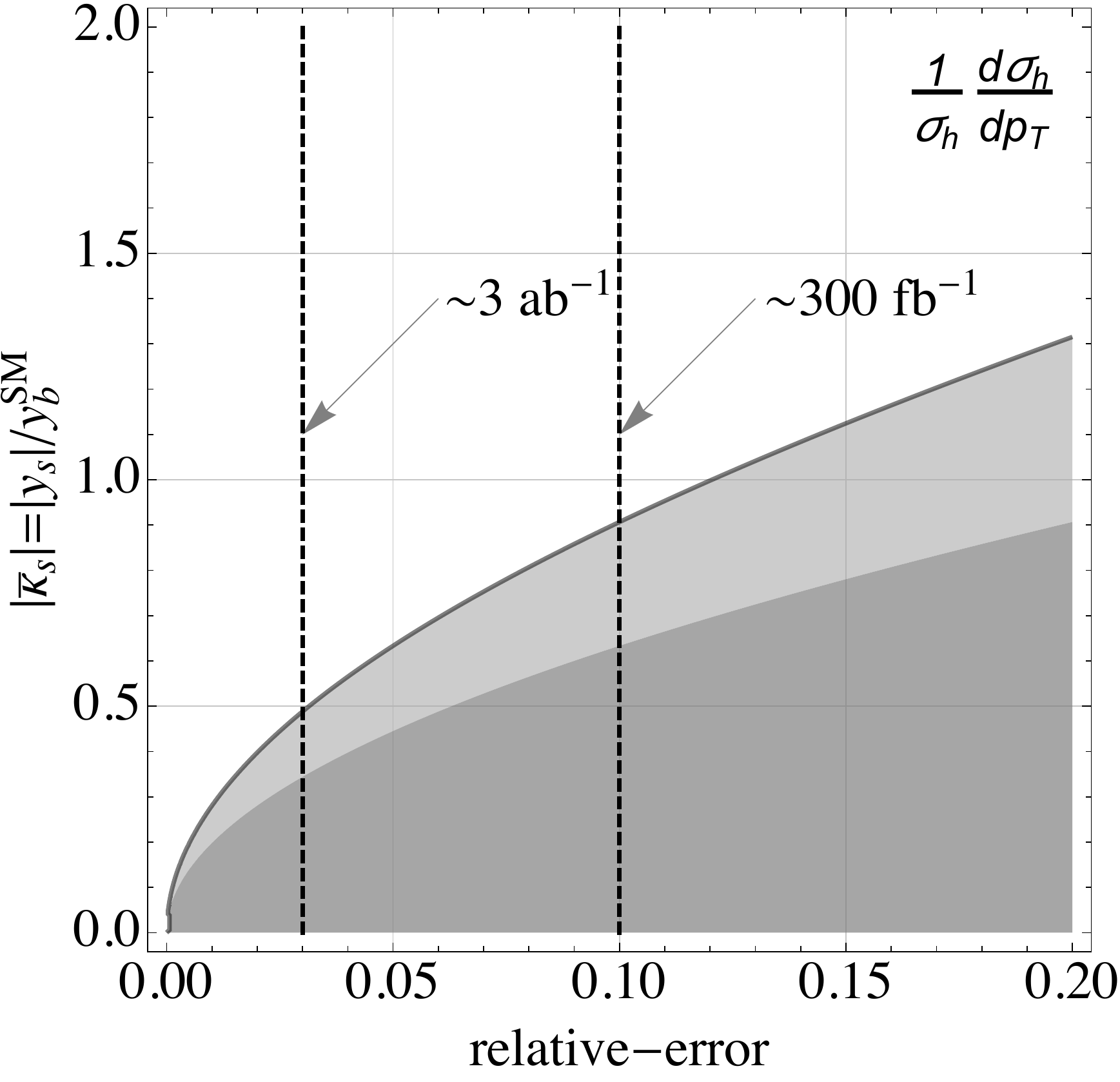}~~~~
\includegraphics[width=0.49\textwidth]{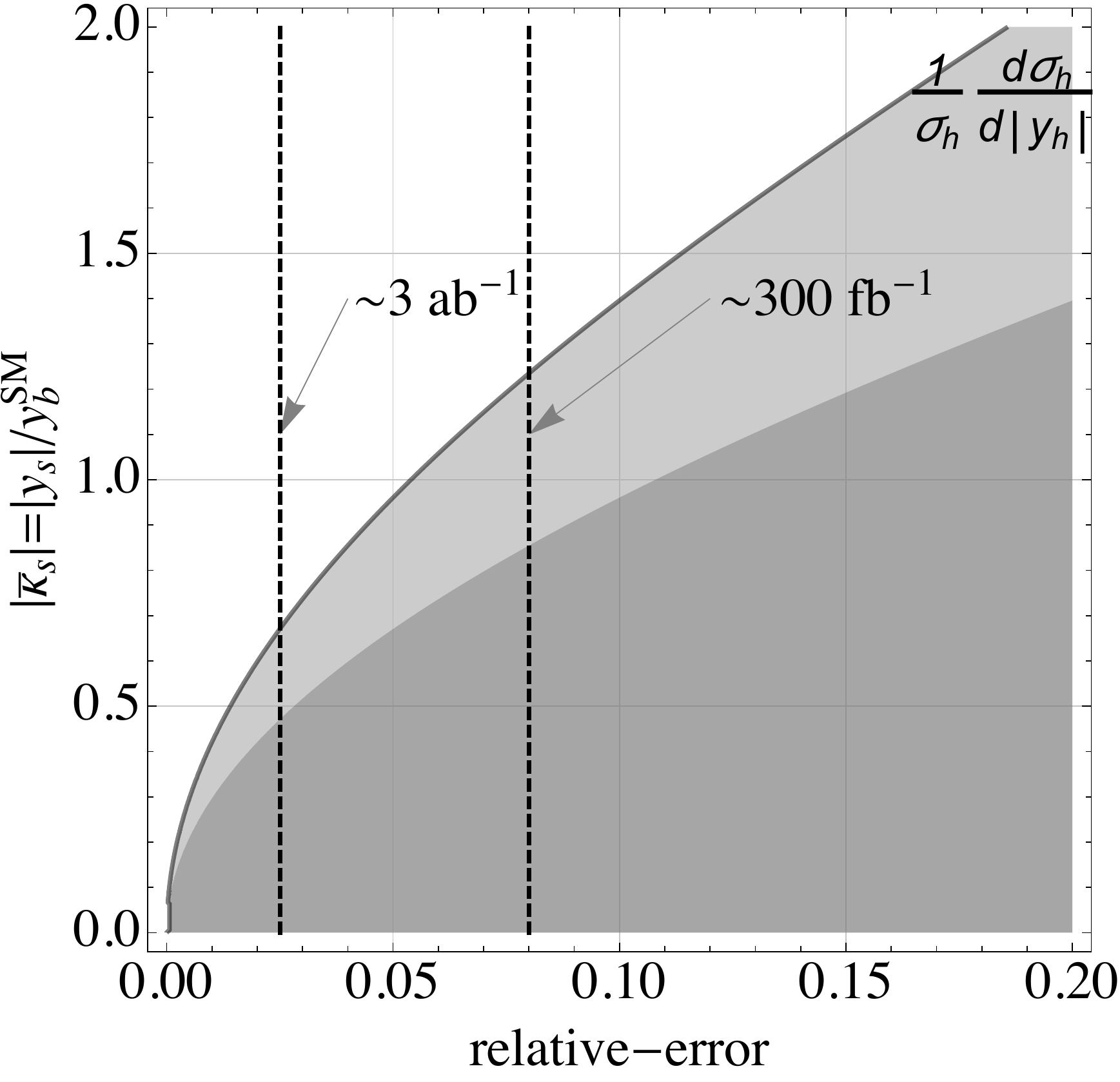}
\caption{Left\,(Right): The dark\,(light) gray region is the $1\sigma$\,$(2\sigma)$ naive projection for probing the strange Yukawa as function of the relative error per bin from the $p_T\,(y)$ distribution for the LHC 13\,TeV assuming bin size of 10\,GeV\,(0.01). { The vertical lines denote expected statistical only errors for integrated luminosities of $3~{\rm ab}^{-1}$ and $300~{\rm fb}^{-1}$.}}
\label{fig:naive13s}
\end{figure}

To estimate the future sensitivity reach for the measurements of $p_T$ and $y_h$ distributions at 13\,TeV LHC, we use the same binings and the covariance matrix as in the 8\,TeV ATLAS measurements but assume perfect agreement between central values of the experimental points and theoretical predictions. We rescale the relative errors in each of the bins by the effective luminosity gain, $(\sigma_h|_{\rm 13 TeV}/\sigma_h|_{\rm 8 TeV})^{-1/2}\cdot({\cal L}_{\rm 13 TeV}/{\cal L}_{\rm 8 TeV})^{-1/2}$. Taking $\sigma_h|_{\rm 13 TeV}/\sigma_h|_{\rm 8 TeV}=2.3$ and ${\cal L}_{\rm 8 TeV}=20.3~{\rm fb}^{-1}$ we get the expected sensitivity from the $p_T$ distribution at 13\,TeV for the luminosity of ${\cal L}_{\rm 13 TeV}=300~{\rm fb}^{-1}$ to be  $\bar \kappa_u< 0.36$ and $\bar \kappa_d < 0.41$ at 95\%\,CL. This should be compared with the expected sensitivity at 8\,TeV, $\bar \kappa_u< 1.0$ and $\bar \kappa_d < 1.2$. (Note that due to a downward fluctuation in the first bin of ATLAS data~\cite{Aad:2015lha}, cf. Fig.~\ref{fig:8TeVpT}~(left), the expected sensitivity is significantly worse than the presently extracted bounds in \eqref{eq:8TeVbounds}.) The expected sensitivities from normalized rapidity distributions are looser,  $\bar \kappa_u< 0.84\,(2.0)$ and $\bar \kappa_d < 1.1\,(3.7)$ for 13\,TeV $300~{\rm fb}^{-1}$\,(8\,TeV $20.3~{\rm fb}^{-1}$). Note that in these rescaling we assumed that the systematic errors will be subdominant, or, equivalently, that they will scale as the statistical errors. Assuming relative error of 5\% in each bin and $p_T$ bins of 10\,GeV we get that $\bar{\kappa}_u<0.27$ and $\bar{\kappa}_d<0.31$. 
This error includes both the systematic and statistical errors. The theoretical error is presently at the level of $\sim 15\%$~\cite{deFlorian:2012mx}~(see e.g. Ref.~\cite{Aad:2015lha}), so that  significant improvement on the theoretical errors were assumed in the above projection, which we find reasonable in light of recent progress on theory~\cite{Li:2016ctv,Melnikov:2016emg,Caola:2016upw}.  To reach the quoted bounds will require a significant amount of data. For instance,  the
statistical error of $5\%$ in the lowest $p_T$ bins would be reached with ${\mathcal O}(2~{\rm ab}^{-1})$ of 13 TeV data. From the rapidity distribution, with bin size of 0.1, we get $\bar{\kappa}_u<0.36$ and $\bar{\kappa}_d<0.47$, see Fig.~\ref{fig:naive13}. 
In Fig.~\ref{fig:naive13s} we also show the projections for how well one can probe the strange Yukawa 
at 13 TeV LHC from $p_T$ (Fig.~\ref{fig:naive13s} left) and $y$ distributions (Fig.~\ref{fig:naive13s} right). We show the reach as a function of relative errors in each bin with 1$\sigma$ ($2\sigma$) exclusions as a dark (light) grey region. We assume $p_T$ bin sizes of 10\,GeV and rapidity of 0.01, respectively.

Flavor non-universality in the down sector is established, if conclusively $\bar \kappa_d< \bar \kappa_b$. ATLAS is projected to be able to put a lower bound on the the bottom Yukawa of $\bar{\kappa}_b > 0.7$~\cite{ATL-PHYS-PUB-2014-011,Perez:2015lra} with ${\cal L}_{\rm 14 TeV}=300~{\rm fb}^{-1}$. Therefore, given the above prospects for probing the down Yukawa from normalized $p_T$ distribution, we expect that there will be indirect evidence for non universality of the Higgs couplings also in the down quark sector.

\section{Conclusions} \label{sec:Conclusions}
Light quark Yukawa couplings can be bounded from normalized $p_T$ and rapidity distributions, $1/\sigma_h\cdot d\sigma_h/dp_T$ and $1/\sigma_h\cdot d\sigma_h/dy_h$, respectively. In these many of the theoretical and experimental errors cancel, while they still retain sensitivity to potential $q\bar q\to h$ fusion. This would make the normalized $p_T$  distribution softer than the SM production through gluon fusion, while the rapidity would become more forward. We presented a reintepretation of the ATLAS measurements of the normalized $p_T$ and rapidity distributions and derived the bounds on up and down quark Yukawa couplings. Owing to a downward fluctuation in the first bin of the distribution one has $y_d^{\rm exp}< y_b^{\rm SM}$ at more than 95 \%CL. With $300\,{\rm fb}^{-1}$ at 13\,TeV LHC, one can furthermore establish non-universality of Higgs couplings to the down quarks, $y_d^{\rm exp}< y_b^{\rm exp}$.

The study performed in this paper is based on LO event generator, which can be improved by using more advanced theoretical tools. For example, it would be useful to compute the rapidity and $p_T$ distribution for $u\bar{u} \rightarrow h$ and $d \bar{d} \rightarrow h$ to higher orders in QCD~\cite{Harlander:2014hya}. Also,  the SM $gg\to h$ inclusive cross section is now known to N3LO level~\cite{Anastasiou:2015ema,Anastasiou:2016cez}. It would be very interesting to push the calculation for rapidity distribution and $p_T$ distribution to N3LO and N3LL, and including the full mass dependence for massive quark loop in the $gg \to hj$ process to NLO (for recent progress, see e.g., Ref.~\cite{Li:2016ctv,Melnikov:2016emg,Caola:2016upw}).

\mysection{Acknowledgements}
We would like to thank Kirill Melnikov and Jesse Thaler for useful discussions.
The work of Y.S. is supported by the U.S. Department of Energy~(DOE) under cooperative research agreement DE-SC-0012567. 
H.X.Z. is supported by the Office of Nuclear Physics of the U.S. DOE under Contract DE-SC-0011090.
J.Z. is supported in part by the U.S. National Science Foundation under CAREER Grant PHY-1151392.

\appendix

 \bibliography{paper_ref}

\end{document}